\documentclass[12pt]{article}
\usepackage{amsmath, graphicx, latexsym,amssymb}

\newtheorem{theorem}{Theorem}[section]
\newtheorem{definition}{Definition}[section]

\newtheorem{lemma}{Lemma}[section]

\title{From gene trees to species trees II: \\
   Species tree inference in the  deep coalescence
model}
\author{Louxin Zhang\\
Departments of Mathematics and
Biochemistry\\
 National University of Singapore\\
 Singapore 119076\\
 Email: matzlx@nus.edu.sg}
\date{}

\begin{document}
\maketitle

\newpage 
\begin{abstract}
 When gene copies are sampled from various species, the resulting gene tree
might disagree with the containing species tree.
The primary causes of gene tree and species tree discord include
 lineage sorting, horizontal gene transfer, and gene
duplication and loss.
Each of these events yields a different parsimony criterion for
inferring the (containing) species tree from gene trees.  With lineage sorting,
 species tree inference is to find the tree minimizing
 extra gene lineages that had to coexist along species
lineages; with gene duplication, it becomes
to find the tree minimizing gene duplications and/or losses.
In this paper, we show the following results:

(i)  The deep coalescence cost is equal to the number of
 gene losses minus two times the gene duplication cost in
the reconciliation  of  a uniquely leaf labeled gene tree and a species
tree. The deep coalescence cost can be computed in linear time
for any arbitrary gene tree and species tree. 

(ii)  The deep coalescence cost is always no less than the 
gene duplication cost in the reconciliation of an arbitrary gene tree and a
species tree.

(iii) Species tree inference by minimizing deep coalescences is NP-hard.

\end{abstract}


\section{Introduction}
 Gene trees are fundamental to molecular systematics.
Traditionally, a gene tree is reconstructed from  DNA sequence
variation at individual genetic loci in a group of species  and is
taken as the phylogenetic tree of the species due to sequencing
technology limitations. However,  when gene copies are sampled from
various species, the resulting gene tree might disagree with the
species tree.  As such, the relationship between gene trees and
species trees has been the focus of many studies (see for example
\cite{Doyle_SysBot_92, Goodman_SysZool_79, Maddison_SysBiol_97,
Page_SysBiol_94, PN88, T89, Wu91}). It has long  been recognized that gene
trees can be used to estimate species divergence time, ancestral
population sizes and
 even the containing species tree although they
 may not accurately
reflect the species tree \cite{Edwards_Evolution_00,
Hey_Genetics_04, Maddison_SysBiol_06}.

The discord of gene trees and the
containing species tree can arise from horizontal gene transfer, lineage
sorting, and gene duplication and loss. The importance of these causes depends
on the considered  genes and species.
Hence, inferring the species tree from gene trees has been investigated
under various parsimony criteria. With lineage sorting (also
called deep coalescence), the problem is to find the tree minimizing
 extra gene lineages that had to coexist along species
lineages \cite{Maddison_SysBiol_97}; with gene duplication, it becomes
to find the tree minimizing gene duplications and/or losses
\cite{Goodman_SysZool_79, Page_SysBiol_94, Guigo_MPE_96,Ronquist_ZoolScr_97}.

\begin{figure}[bth!]
\begin{center}
\includegraphics[width=0.6\textwidth]{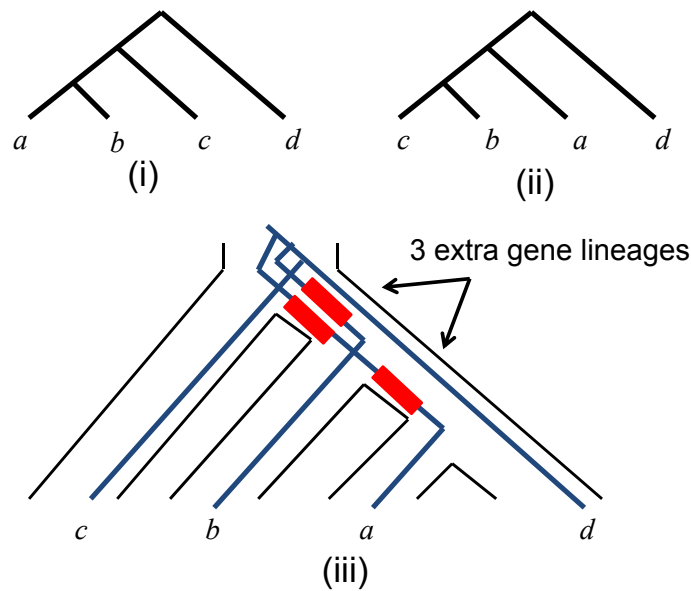}
\end{center}
\caption{(i) A gene tree. (ii) A species tree. (iii)
  The reconciliation of  the gene tree in (i) into the species tree in (ii) has
deep coalescence cost 3.
\label{DCExample}
}
\end{figure}

Inferring the species tree from a set of gene trees has often been  studied
under the gene duplication cost \cite{Bansal_TCBB_08, Chauve_JCB_08,
Chen_JCB_00, Durand_JCB_06, Eulenstein_JCB_98, Hallett_Recomb00, Hadas_JCB_09,
Luo_TCBB_10, Roth_JExpZool_07,Zh97} until very recently.
In a seminal work \cite{Maddison_SysBiol_97}, Maddison addressed lineage sorting
in the framework of coalescence theory. 
Coalescence theory is an active branch of population genetics concerned with
tracing the genealogical history of a present-day gene copy. For a gene sampled
from two individuals, one may ask: How deep in time do these two lineages
coalesce? Hence, the depth of this coalescence is  a measure of the 
relationship between two sampled gene copies. The more deep in time coalescence
occurs, the more distantly related they are.  Maddison
proposed to use the total number of
``extra" gene lineages that fails to coalesce on a species tree
 to measure the inconsistence of a gene tree and species tree,
called {\it deep coalescence
cost}. For the gene tree and species tree shown in
Figure~\ref{DCExample},  there are three gene lineages
 on a branch and two gene lineages on  another branch
that fail to coalesce, giving the deep coalescence cost of 3.
 Since coalescence
theory provides the probability that a gene tree would exist in a
species tree, it allows the inference problem to be studied in
explicit statistical framework \cite{Degan_Evol_05,
Rosenberg_TheorPopBiol_2002}. This seems to give the deep
coalescence model an advantage over the other models.

The paper is a sequel of \cite{Ma_SIAMComput_01}, which studies the
complexity and algorithmic issues of inferring the species tree from
a set of gene trees with the gene duplication/loss cost.  In this
work, we present a relationship of  the deep coalescence cost, the
duplication cost, and the number of gene losses. Although deep
coalescence and gene duplication are two different mechanisms
responsible for the discord of gene trees and species trees, this
relationship suggests that the deep coalescence cost and the
duplication cost are closely related to each other as a similarity
measure of trees. We further show that inferring species tree from
gene trees is also NP-hard  by minimizing the deep coalescence cost.

\section{Basic definitions and notations}

In this section, we shall introduce basic definitions and notations on
gene duplication, gene loss and deep coalescence that are used in
 the following sections.

\subsection{Species trees and gene trees}

For a set of $n$ taxa, their evolutionary history is modeled as a rooted, full
binary tree with $n$ leaves in which leaves are labeled with taxa,
 representing the labeling taxa,
and internal nodes are unlabeled.   Here, the 'fullness' means that
each internal node has exactly two children. Such a tree is called
{\it species tree}. In a species tree, each unlabeled internal node
is considered as a {\it taxon family} which include as its members the
subordinate species represented by the leaves below it.
Thus, the evolutionary relation ``$m$ is a descendant of $n$" is expressed
using the set-theoretic notation as ``$m\subset n$".
  We also call an internal node  an {\it ancestor} of the species below it.

The model for gene relationship is also a rooted, full binary tree
with leaves representing genes, called a {\it gene tree}. Usually, a
gene tree is reconstructed from a collection of gene  family members
sampled from the considered species. We label the gene copies by the
species from which they are sampled.
 Thus, leaf labels may not be unique in a gene tree as two or more  gene
copies might  be found in a species.
An internal node $g$ corresponds to a multiset of leaf labels.

Finally, for a species or  gene tree $T$, we use $L(T)$ to denote the set
of leaf labels of it. For an internal node $t$ in $T$, $a(t)$ and $b(t)$ are
used to denote its two children.

\subsection{Gene duplication}

Let $G$ be a gene tree and $S$ a species tree such that
$L(G)\subseteq L(S)$. For any nodes  $s', s''$ in $S$, the {\it least
common ancestor} of $s'$ and $s''$ is defined to be the smallest node
$s$ in $S$ such that $s', s''\subseteq s$, which is denoted by
$\mbox{lca}(s', s'')$.  To reconcile the gene tree  $G$ and the species tree
 $S$, each node $g$ of $G$ is mapped
to a unique node $M(g)$ in $S$ as
\begin{eqnarray*}
 M(g)=\left\{ \begin{array}{ll}
              g, & g\in L(G)\\
            \mbox{lca}\left(M(a(g)), M(b(g))\right), & g\not\in L(G).
          \end{array}
        \right.
\end{eqnarray*}
This mapping $M$ was first considered in \cite{Goodman_SysZool_79} and then
formulated in
\cite{Page_SysBiol_94}. We call $M$ the $\mbox{lca}$ mapping or reconciliation of $G$ in $S$.
Obviously, if $g'\subset g$, $M(g') \subseteq M(g)$.

\begin{definition}
 Let $g$ be an internal node of $G$. If
 $M(c(g))=M(g)$ for some child $c(g)$ of $g$, then
we say that  a duplication occurs at $M(g)$ (or more exactly
in the lineage entering $M(g)$) in $S$.
\end{definition}

The total number of duplications arising in the lca reconciliation  $G$ in  $S$
 is proposed to measure the discord
of the gene tree and species tree  and is called the
{\it duplication cost}. We use $c_{dup}(G, S)$ to denote the duplication cost
for $G$ and $S$. Note that the duplication cost is not symmetric.

\subsection{Gene loss}

A subset $A$ of (internal and/or leaf) nodes of a species tree $S$ is
 incompatible  if $x\cap y=\phi$ for any $x, y\in A$. For an incompatible
 subset
$A$ in $S$, the restriction of $S$ on $A$ is the smallest subtree
of $S$ containing $A$ as its leaf set, denoted by $R_S(A)$.
It is easy to see that
the root of $R_S(A)$ is the least common ancestor of the nodes from $A$.
The homomorphic subtree $S|_A$ of $S$ induced by $A$ is a tree obtained
from $R_S(A)$ by contracting all degree-2 nodes except for the root
of $R_S(A)$.

Let $G$ be a gene tree such that  $L(G)\subseteq L(S)$.
 $S|_{L(G)}$ is well defined.
To reconcile $G$ and $S$ in this general case,
we consider the the lca mapping $M$ from $G$ to $S|_{L(G)}$.
For any two nodes  $s$ and $s'$  of $S|_{L(G)}$ such that $s\subset s'$,
 we define
$$
 d(s, s')=|\{ h\in S|_{L(G)} | s \subset h \subset s'\}|.
$$
That is, $d(s, s')$ is the number of nodes on the path from $s'$ to $s$.

Recall that $a(g)$ and $b(g)$ denote the children of $g$. The number of
losses $l_g$ associated to $g$ is defined as
\begin{eqnarray*}
l_g=\left\{ \begin{array}{ll}
  0, & \mbox{  if }  M(g)=M(a(g))=M(b(g)),\\
d(M(a(g)), M(g)) +1, & \mbox{ if }  M(a(g))\subset M(g)= M(b(g)),\\
\sum_{h=a(g),b(g)}d(M(h), M(g)), & \mbox{ if }
M(a(g)), M(b(g))\subset M(g).
\end{array}
\right.
\end{eqnarray*}
 This definition of $l_g$ is a generalization of  the loss cost given
in \cite{Guigo_MPE_96}. When $L(G)=L(S)$, our definition  is then
identical to the one given in \cite{Guigo_MPE_96}.

The {\it gene loss cost} in the reconciliation of $G$ in $S$
is defined as the total number of losses $\sum_{g\in G} l_g$.
We denoted this gene loss cost for $G$ and $S$  by $c_{loss}(G, S)$.

\subsection{Deep coalescence}.

Let $G$ be a gene tree and $S$ a species tree such that
$L(G)=L(S)$.  Under the lca mapping $M: G\rightarrow S$,
if a branch $e$ of $S$ is on the $k$ paths from $M(g_i)$ to $M(c(g_i))$,
$g_i\in G$ ($1\leq i\leq k$),  then we say that
there are $k-1$ `extra' lineages on $e$ failing to coalesce on  $e$.
The {\it deep coalescence}(DC) {\it cost} is defined as the total number of the
`extra' lineages on all branches of $S$ in the reconciliation $M$ of $G$ in $S$
(see \cite{Maddison_SysBiol_97}), which is denoted by $c_{dc}(G, S)$.
Note that the concept of deep coalescence is meaningful only if 
$S$ has 2 or not leaves. We assume this throughout the paper.

In general, if $L(G)\subset L(S)$, the deep coalescence cost
$c_{dc}(G, S)$ is defined as $t_{dc}(G, S|_{L(G)})$, where
$S_{L(G)}$ is the homomorphic subtree of $S$ induced by $L(G)$. Such
a generalization will be used in the study of inferring the species
tree from a set of gene trees

\section{A equation of the duplication  and  DC costs}

 We have seen that  deep coalescences, gene losses and duplications
are inferred through the gene tree/speceis tree reconciliation.
Actually, they are indeed closely related through a simple equation.

\begin{definition}
 Let $G$ be a gene tree and $S$ a species tree such that $L(G)\subseteq L(S)$.
Under the lca  mapping ${M}: G\rightarrow S$,  an internal node  $g\in G$ is of
 \begin{itemize}
  \item  type-1 if $M(g') \subset M(g)$ for each child $g'$ of $g$;
  \item  type-2 if there exists a unique child $g'$ such that $M(g') =
M(g)$;
  \item type-3  if $M(g')=M(g)$ for  each child $g'$ of $g$.
 \end{itemize}
\end{definition}

 Note that type-2 or type-3 internal nodes correspond one-to-one with duplication events.

\begin{theorem}
\label{thm31}
Let $G$ be a uniquely leaf-labeled gene tree and $S$ a species tree such
that $L(G)=L(S)$. Then,
  $$ c_{dc}(G, S) = c_{loss}(G, S) - 2c_{dup}(G, S).$$
\end{theorem}
{\bf Proof.} Let $G$ and $S$ have $n$ leaves.
Assume that there are $k_1$ type-1 internal nodes
 $$ g_{11}, g_{12}, \ldots,  g_{1k_1},$$
$k_2$ type-2 internal nodes
 $$ g_{21}, g_{22}, \ldots, g_{2k_2}, $$
 and $k_3$ type-3  internal nodes
$$ g_{31}, g_{32}, \ldots, g_{3k_3}$$
in $G$ under the lca  mapping $M: G\rightarrow S$, respectively.
Since $G$ is a full binary tree with $n$ leaves, $G$
has   $n-1$ internal nodes and hence
\begin{eqnarray}
\label{node_partition}
 k_1+k_2+k_3 = n-1.
\end{eqnarray}
Additionally,  type-2 and type-3 nodes correspond one-to-one with
duplication events,
\begin{eqnarray}
\label{dup_cost}
  c_{dup}(G, S)= k_2+k_3.
\end{eqnarray}
For simplicity, we assume that
 $g'$ and $g''$ are the children of $g$ for each type-1 internal node $g$;
we also assume  that $a(g)$ is the
unique child such that $M\left(a(g)\right)\subset M(g)$
for each type-2 node $g$.
Since we use  $d\left(M(h), M(g)\right)$ to denote the number of
nodes on the path from $M(g)$ to $M(h)$ for a node $g$ and its child $h$,
the number of lineages contained in the path is
$d\left(M(h), M(g)\right)+1$.  Therefore,
by Eqn.~(\ref{node_partition}) and~(\ref{dup_cost}) and  the fact that $|E(S)|=2n-2$,
\begin{eqnarray*}
  c_{dc}(G, S)& =& \sum ^{k_1} _{j=1}  \left\{ \left[d\left(M(g'_{1j}), M(g_{1j})\right)+1\right] +
   \left[d\left(M(g''_{1j}), M(g_{1j})\right)+1\right]\right\}\\
 & & + \sum ^{k_2} _{j=1} \left[d\left(M(a(g_{1j})), M(g_{1j})\right)+1\right] -|E(S)|\\
  & = & c_{loss}(G, S) + 2k_1 - (2n-2)\\
  & = & c_{loss}(G, S) -2(k_2+k_3) \\
  &=  &  c_{loss} (G, S) - 2c_{dup}(G, S).
\end{eqnarray*}
This concludes the proof. $\Box$
\vspace{1em}


\noindent {\bf Remarks}. (1) Following the proof of the equation in the
above theorem, one can easily see that for an arbitrary gene tree
$G$ in which there may be two or more gene copies are from the same
species and a species tree $S$ such that $L(G)=L(S)$,
$$c_{dc}(G, S)= c_{loss}(G, S) - 2c_{dup}(G, S) + (no. of genes)-(no. of
species).$$

(2) Since the number of gene  duplications and lossed can be calculated in
linear time \cite{Zh97, Ma_SIAMComput_01}, the first remark implies that
the deep coalescence cost can also be computed in linear time. 
\vspace{1em}

 By Thm~\ref{thm31}, $c_{dc}(G, S)\leq c_{loss}(G, S)$ for a species tree $S$
and a uniquely leaf labeled gene tree $G$. Now we show that it is bounded below
by the duplication cost for any arbitrary gene tree.

\begin{figure}
\begin{center}
\includegraphics[width=0.6\textwidth]{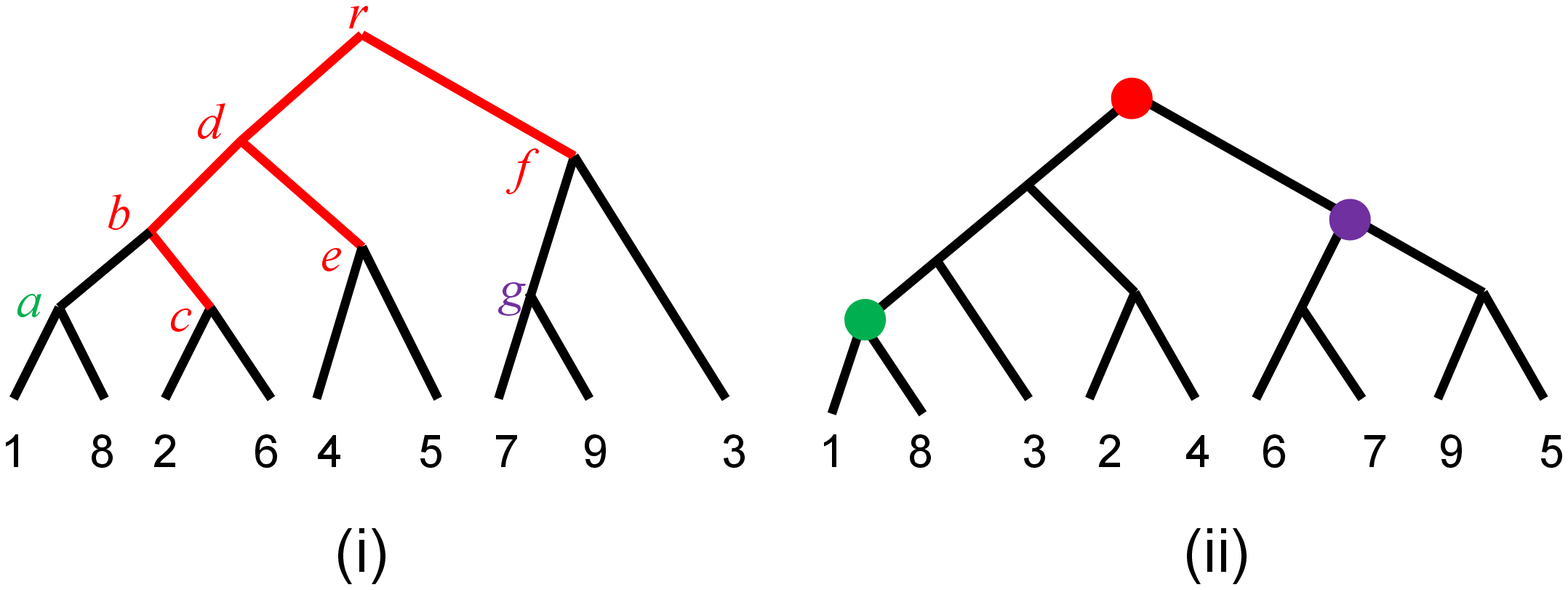}
\end{center}
\caption{(i) A gene tree. (ii) A species tree. In the lca  reconciliation $M$
  of the gene tree in the species tree,   $a$ is mapped to the green node,
$b, c, d, e, f$ and $r$ to the red node, and $g$ to the purple
node. The nodes $b, c, d, e, f, r$ form a subtree of the gene tree.}
\label{GeneTreePartition}
\end{figure}

\begin{theorem}
\label{DC_DupLoss_Thm}
 Let $G$ be a uniquely leaf-labeled gene tree
and $S$  a species tree such that $L(G)=L(S)$. Then,
    $c_{dc}(G, S) \geq c_{dup}(G, S)$.
\end{theorem}
{\bf Proof.}
 Denote the image node set  of the lca  mapping $M$ by $M(G)$, which is a subset
of nodes in the species tree $S$.
For any internal node $s\in M(G)$,
we use $M^{-1}(s)$ to denote all internal nodes $g$ of the gene tree
 that are mapped
to $s$ under $M$. For any nodes $x$ and a  descendant $y$ of $x$ in the gene tree $G$,
 if $M(x)=M(y)=s$, then $M(g)=s$ for each node in the path from $x$ to
$y$.  Since $G$ is uniquely leaf labeled,
all internal nodes in $M^{-1}(s)$ form a rooted subtree of $G$,
denoted by $T^{-1}(s)$,  as illustrated in Figure~\ref{GeneTreePartition}.

$T^{-1}(s)$ is not a full binary tree in general. In particular,
its root might has degree 1.
Let $n'_{s}, n''_s, n'''_s$  denote the number of non-root
degree-1, degree-2 and
degree-3 nodes in the subtree $T^{-1}(s)$, respectively.
Assume that $T^{-1}(s)$ has  two  or more nodes.
Then, by definition,  the root of $T^{-1}(s)$ corresponds with a gene duplication
in the reconciliation of $G$ and $S$;
  each degree-2 or degree-3 node of $T^{-1}(s)$ also corresponds with a
gene duplication. Therefore, there are $n''_s+n'''_s+1$
 duplication events at $s$.  We now consider two cases.

 Case 1. The root of $T^{-1}(s)$ has degree 1.   Then
  $T^{-1}(s)$  has  $n'''_s+1$ leaves, that is $n'_s=n'''_s+1$.
For  each leaf of $T^{-1}(s)$, it has two children that are mapped to
a node  below $s$ in the species tree $S$;
each non-root degree-2 node has exactly one child that is
mapped to a node below $s$ and so is the root since it has degree 1.
Thus,
there are $2\left(n'''_s+1\right) + n''_s + 1$
 image paths that contain  one of  the two lineages from $s$ to one of its children.

Case 2.  The root has degree 2.
 In this case,  $T^{-1}(s)$ has $n'''_s+2$ leaves and
there are $2\left(n'''_s+2\right)+n''_s$ image paths that contain
 one of the two lineages from $s$ to one of  its children.

By distributing the DC and duplication costs to each image node $s$ in
$M(G)$, we obtain
that
 \begin{eqnarray}
  &&   c_{dc}(G, S)\\
 & \geq  & \sum_{s\in M(G): |T^{-1}(s)|>1}
    (\mbox{the no. of extra gene lineages on the branches leaving $s$})
\nonumber\\
   & \geq &  \sum_{s\in M(G): |T^{-1}(s)|>1}  \left(2n'''_s+n''_s+1\right)
\nonumber \\
   & \geq &  \sum_{s\in M(G): |T^{-1}(s)|>1}  \left(n'''_s+n''_s+1\right)
\nonumber \\
   & =& c_{dup}(G, S). \label{lowerBound_est}
\end{eqnarray}
This finishes the proof. $\Box$
\vspace{1em}

\noindent {\bf Remark} The fact $c_{dc}(G, S)\geq c_{dup}(G, S)$ holds
even for arbitrary gene trees in which 2 or more leaves with  the same label,
which represent genes sampled from the same species.
In the  general case, $T^{-s}$ might be a forest -- a union of rooted trees.
However, the estimation  (\ref{lowerBound_est}) in the proof is still
valid if the sum is over all the subtrees that are mapped to a node in the
species tree, i.e.  $T^{-s}$ is replaced by a subtree of each resulting forest.

\section{The NP-harness of the species tree problem in the DC cost}

Parsimony criterion is often used for inference  in
biology. Hence, inferring  species tree from  a set of gene trees is
formulated as the following algorithmic problem
\vspace{1em}

\noindent {\bf Species Tree Problem}\\
{\sc Input}: A set of gene trees $G_i$, $1\leq i\leq n$.\\
{\sc Solution}: A species tree $S$ that minimizes the total cost
    $\sum_{i} c(G_i, S)$, where $c( , )$ is a cost function.
\vspace{1em}

It is proved that the species tree problem is NP-hard for the duplication
and/or loss cost in \cite{Ma_SIAMComput_01}. In this section, we prove
the following theorem.

\begin{theorem}
The species tree problem is NP-hard under the DC cost.
\end{theorem}
{\bf Proof.} Given a gene tree $G$ and a species tree $S$,
the DC cost $c_{dc}(G, S)$ can be
computed in polynomial time since gene duplications and
losses can be  counted in linear time \cite{Zh97}.
Therefore,  the species tree problem is in NP.

 To prove its NP-hardness, we reduce the Maximum Cut problem to the
 decision version of the species tree problem. Given an instance
 graph ${\cal G}=(V, E)$ and
a positive integer $I$,  the Maximum
 Cut problem is to partition the node set $V$ into two disjoint
 subsets $V_1$ and
 $V_2$ such that there are at least $I$  edges from $E$ that have one
 endpoint in $V_1$ and one endpoint in $V_2$.  Assume
 that $V=\{v_1, v_2, \cdots, v_n\}$ and $|E|$ denotes the number of
 edges from $E$, where $n>3$. We construct a corresponding instance of the
 species tree problem as follows.

\begin{figure}[th!]
\begin{center}
\includegraphics[width=0.8\textwidth]{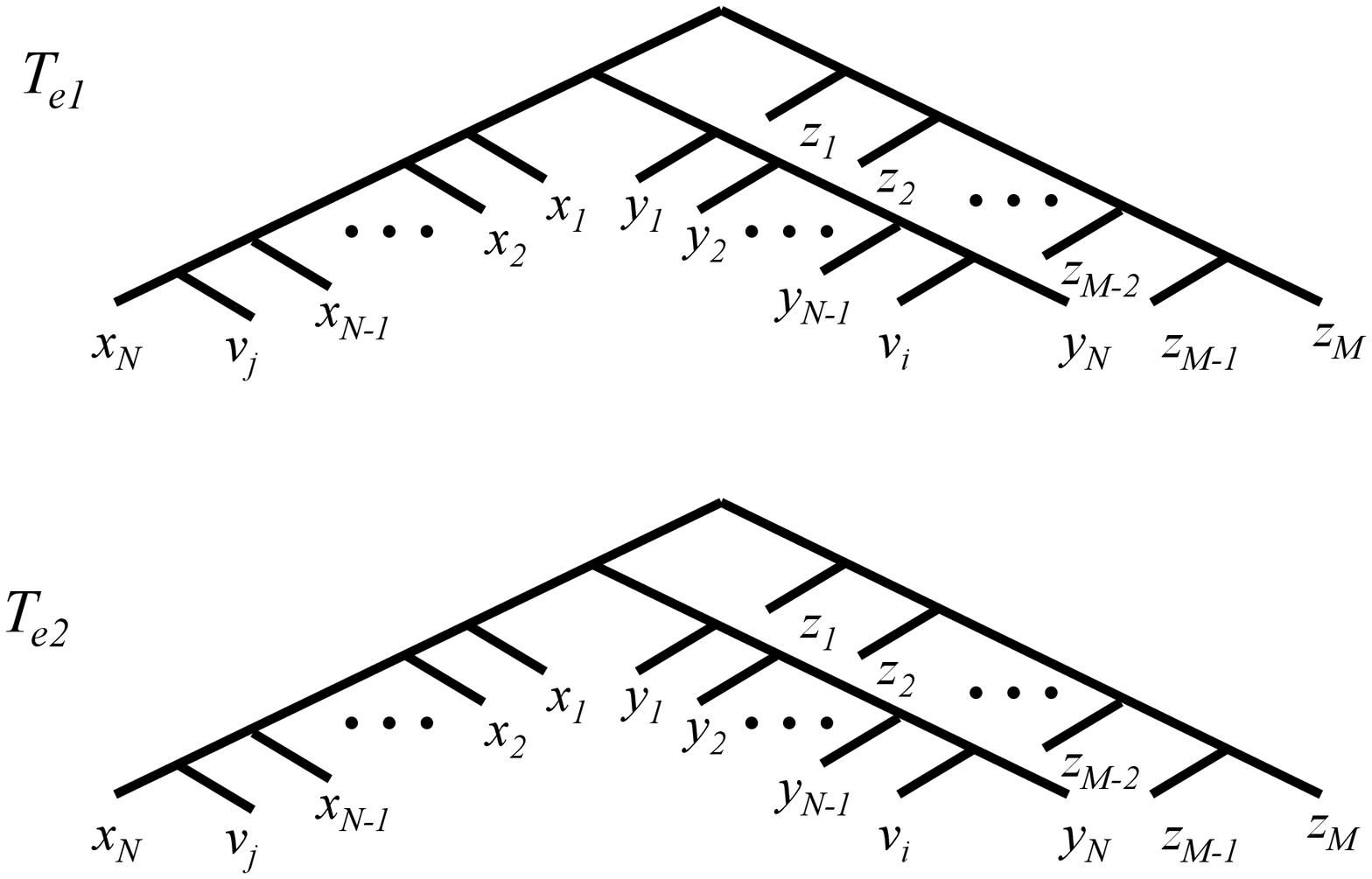}
\end{center}
\caption{Gene trees defined for each edge $e=(v_i, v_j)$.}
\label{EdgeTrees}
\end{figure}

 Choose  $N>n^2$ and $M\geq n^2N(N+1)+ |E|$.
 For each node $v_i$ ($1\leq i\leq n$), we introduce a label with the same
name $v_i$. We
 also introduce $2N+M$ extra labels $x_i, y_i$, $1\leq i\leq N$ and $z_j$, $1\leq j\leq M$.
 For each edge $e=(v_i, v_j) \in E$, we define two gene trees $T_{e1}$ and
 $T_{e2}$ as shown in Figure~\ref{EdgeTrees}.
These two trees are same except that  the leaf labels $v_i$ and  $v_{j}$ are
swapped.

\begin{figure}[tb!]
\begin{center}
\includegraphics[width=0.7\textwidth]{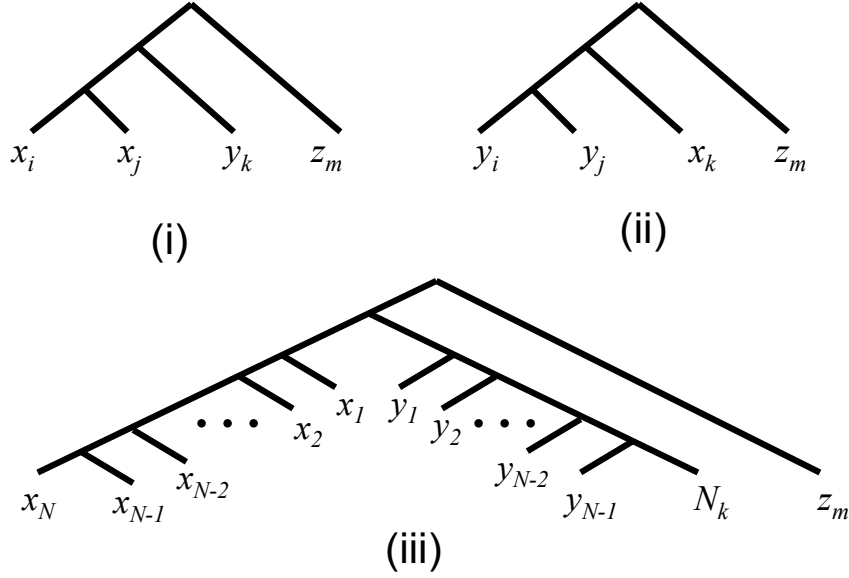}
\end{center}
\caption{`Structural' gene trees.}
\label{Gaget_Trees}
\end{figure}

 Let the trees shown in Figure~\ref{Gaget_Trees} (i)-(iii) be written as
 $L[x_i, x_j, y_k, z_l]$, $L[y_i, y_j, x_k, z_l]$ and $F[\{x_i\},
 \{y_i\}, z_l]$, respectively.
Besides the `edge' gene trees $T_{e1}$ and $T_{e2}$ ($e\in E)$, the set $\cal A$ of gene trees
in the instance of the problem to be defined also contains
 \begin{eqnarray*}
    G_{(i, j, k, m)} =  L[x_i, x_j, y_k, z_m],
   & 1\leq i<j \leq N,\;1\leq k\leq N, \; 1\leq m\leq M, &\\
   G'_{(i, j, k, m)} = L[y_i, y_j, x_k, z_m],
   & 1\leq i<j \leq N,\;1\leq k\leq N, \; 1\leq m\leq M, & \\
   G''_{m}=F[\{x_i\},\{y_i\}, z_m],& 1\leq m\leq M. &
 \end{eqnarray*}
These three classes of gene trees are introduced to restrict the topology of
the optimal species tree for the defined instance of the problem. Hence, we
call them `structural' gene trees.
 The NP-completeness of the decision version of the species tree problem
follows from the following two lemmas.

\begin{figure}[tb!]
\begin{center}
\includegraphics[width=0.8\textwidth]{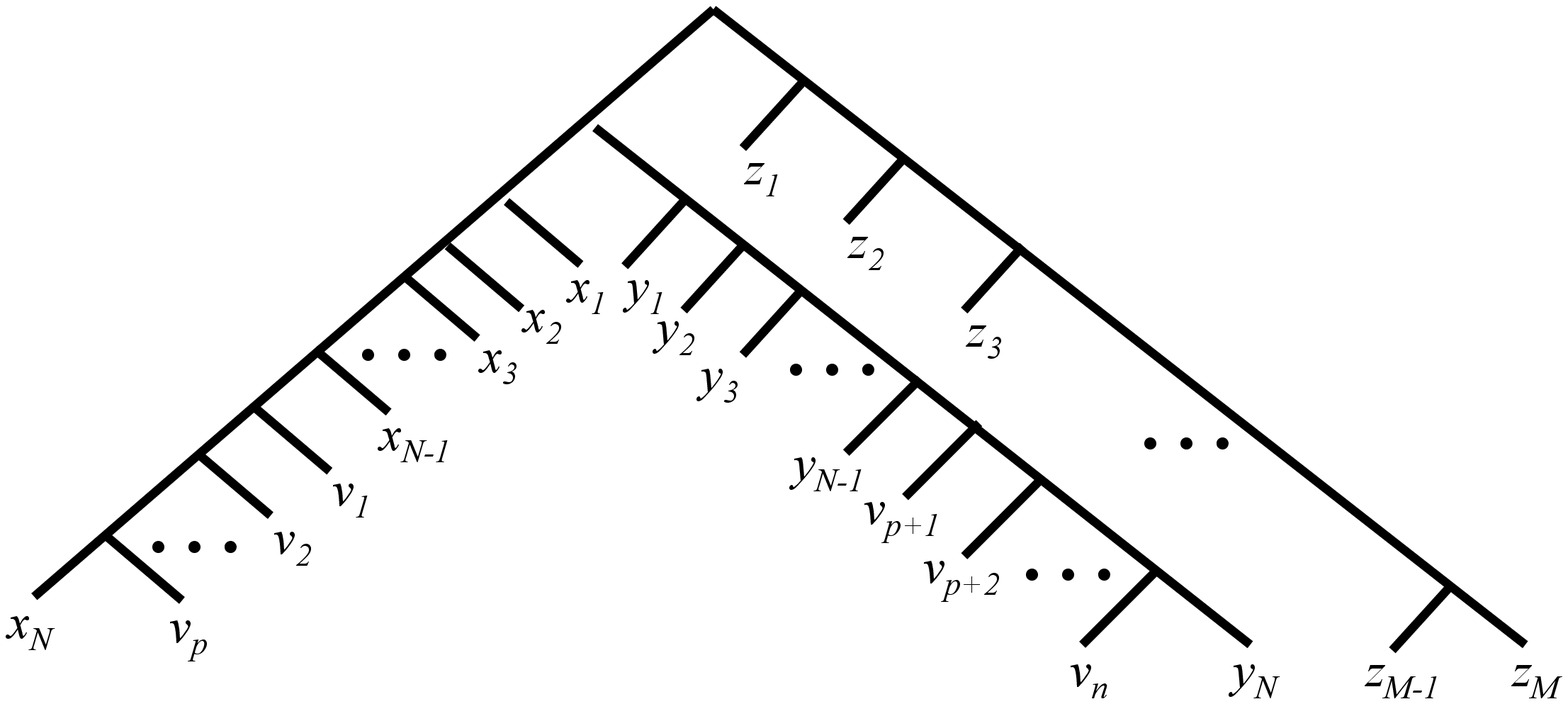}
\end{center}
\caption{Species tree $S_{\cal G}$ defined in Lemma~\ref{lemma1}.}
\label{SpeciesTree}
\end{figure}

 \begin{lemma}
\label{lemma1}
   If the graph $\cal G$ has a cut of $d$ edges, there is a
   species tree $S_{\cal G}$ having the DC cost
   $$c_{dc}({\cal A}, S_{\cal G})= N(N+1) |E| + |E| -d.$$
 \end{lemma}
 {\bf Proof.} Assume that the node set $V$ of the graph $\cal G$ divides
 into $V_1=\{v_1,v_2, \cdots, v_p\}$ and $V_2=\{v_{p+1},
 v_{p+2}, \cdots, v_n\}$ such that there are exactly $d$ edges having one
 endpoint in $V_1$ and one endpoint in $V_{2}$.  We define a species
 tree $S_{\cal G}$ as shown in Figure~\ref{SpeciesTree}.

  First, we observe that
 \begin{eqnarray*}
   c_{dc}(G_{(i, j, k, m)}, S_{\cal G})=0,&
   c_{dc}(G'_{(i, j, k, m)}, S_{\cal G})=0, &
   c_{dc}(G''_{m}, S_{\cal G})=0,
 \end{eqnarray*}
for each possible $i, j,k, m$.

 Consider a non-cut edge $e=(v_i, v_j)$ ($i<j$). Since
$L(T_{e1})=L(T_{e2})\subset L(S_{\cal G})$,
$c_{dc}(T_{e1}, S_{\cal G})=c_{dc}(T_{e1}, S_{\cal G}|_{L(T_{e1})})$ and
$c_{dc}(T_{e2}, S_{\cal G})=c_{dc}(T_{e2}, S_{\cal G}|_{L(T_{e2})})$.
  If   $v_i, v_j\in V_1$, we have that
    \begin{eqnarray*}
       c_{dc}(T_{e1}, S_{\cal G})= \frac{1}{2}N(N-1)+ N + 1,\;\;
       c_{dc}(T_{e2}, S_{\cal G})= \frac{1}{2}N(N-1)+ N.
    \end{eqnarray*}
 Symmetrically, if $v_i, v_j\in V_2$, we have that
  \begin{eqnarray*}
       c_{dc}(T_{e1}, S_{\cal G})= \frac{1}{2}N(N-1) + N ,\;\;
       c_{dc}(T_{e2}, S_{\cal G})= \frac{1}{2}N(N-1)+ N+1.
    \end{eqnarray*}

  For each cut edge $e=(v_i, v_j)$ ($i<j$) with one endpoint in $V_1$, say
$v_i\in V_1$, and another  in $V_2$, we
have that
    \begin{eqnarray}
       c_{dc}(T_{e1}, S_{\cal G})=0,&
       c_{dc}(T_{e2}, S_{\cal G})= N(N+1).
    \end{eqnarray}
Therefore, we have
  $$c_{dc}({\cal A}, S_{\cal G})= N(N+1)|E|+|E|-d.$$
This finishes the proof of the lemma. $\Box$

\begin{figure}[tb!]
\begin{center}
\includegraphics[width=0.7\textwidth]{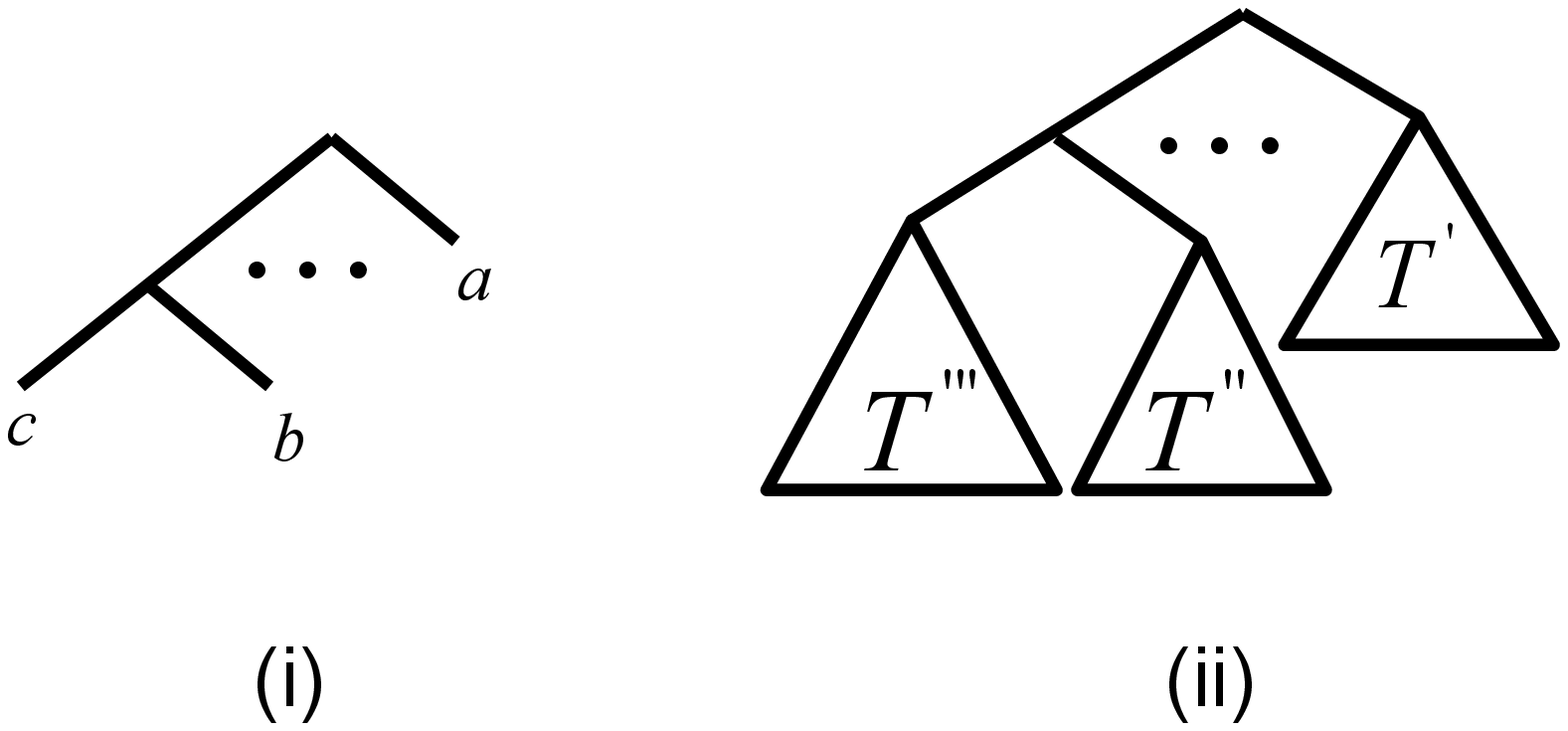}
\end{center}
\caption{(i) Line tree $\mbox{LTree}[a, \ldots, b, c]$. (ii)
The resulting tree $\mbox{LTree}[T',
\ldots, T'', T''']$ after replacing each leaf with a tree in a line tree.}
\label{LineTree}
\end{figure}

 \begin{lemma}
\label{lemma2}
    If there is a species tree $S$ having the DC cost
      $c_{dc}({\cal A}, S)=N(N+1)|E| + t$,
         then the graph $\cal G$ has a cut of at least
      $|E|-t$ edges.
 \end{lemma}
{\bf Proof.}  
If $t> |E|$, the fact is trivial. Hence, without loss of generality,
 we may assume that  $t\leq |E|$.
 Here,   we use $\mbox{LTree}[a,  \ldots, b,  c]$ to denote the line tree
with leaves labeled by $a$, $b$, $\ldots$, $c$, respectively, as shown
in Figure~\ref{LineTree} (i).
Note that the leaf $a$ is a child of the root in  $\mbox{LTree}[a,  \ldots, b,
c]$.
 For a set of trees $T'$, $T''$, $\ldots$,  $T'''$,
we use
 $$\mbox{LTree}[T', \ldots, T'',  T''']$$
 to denote the tree
obtained by replacing each leaf by a corresponding subtree in
 $\mbox{LTree}[a, \ldots, b,  c]$ as shown in Figure~\ref{LineTree} (ii).

   Let $B$ be a subset of leaves in the species tree $S$ and  the least common
ancestor of the leaves from $B$  be $r_{B}$ in $S$.   Recall that the
   homomorphic subtree $S|_B$ of $S$ induced by $B$ is the tree
   obtained from $S$ by removing all the nodes and edges that are
   not on a path from $r_B$  to a leaf from $B$ and then
   contracting all the degree-2 node except for the root $r_{B}$.
  For example, for $S_{\cal G}$ defined in Lemma\ref{lemma1},
$S_{\cal G}|_{\{x_1, x_2, y_1\}}=\mbox{LTree}[y_1, x_1, x_2]$.

   Set
   \begin{eqnarray}
    &  U=\{x_1, x_2, \cdots, x_N\}\cup \{ y_1, y_2, \cdots, y_{N}\}
     \cup \{ v_1, v_2, \cdot, v_n\}; & \nonumber \\
      & Z=\{z_1, z_{2}, \cdots, z_M\}.&
     \end{eqnarray}
   By replacing the children of a two-leaf root tree with $S|_{U}$ and
   $S|_{Z}$, we obtain a species tree $S'=\mbox{LTree}[S|_{U}, S|_{Z}]$ from S.
 First, $S'$ has the following property.
\vspace{1em}

\noindent   {\bf Fact 1}  $c_{dc}({\cal A}, S') \leq c_{dc}({\cal A}, S)=N(N+1)|E| + t$.

\noindent {\bf Proof.} For each gene tree $T=T_{e1}$ or $T_{e2}$, we use $f$ and $f'$ to
denote the lca  mappings from $T$ to $S$ and $S'$, respectively.
  For each edge $e=(u_1, u_2)$ in the spanning subtree over $z_i$s in $T$,
by the definition of $S|_{Z}$, $f(u_1)=f(u_2)$ if and only if
$f'(u_1)=f'(u_2)$ and  $d(f'(u_1), f'(u_2)) \leq d(f(u_1), f(u_2))$.
 For each edge in the spanning subtree over $x_i$s, $v_i$, $v_j$ and $y_i$s,
the same property holds. 
But, the edges incident to the root of $T$ may not satisfy the property 
discussed above. Let $r$ be the root of $T$. Assume that $a(r)$ 
is the left child of $r$, which is the least common ancester of $x_i$s and 
$y_i$s,  and $b(r)$  the right child of $r$.
It is possible that $f(r)=f(a(r))$ and/or $f(r)=f(b(r))$. 
However, $f'(r)=r'$,  $f'(a(r))=a(r')$ and $f'(b(r))=b(r')$,
where $r'$ is the root of $S'$, $a(r')$ and $b(r')$ 
 the root of $S|_{U}$ and $S|_{Z}$ respectively.
 Since no other lineages fail to coalesce with 
$(r, a(r))$  on $(r', a(r')$ and with $(r, b(r))$ on
 $(r', b(r'))$ respectively, these two edges
does not affect the deep coalescence cost. 
Thus,  $c_{dc}(T, S')\leq c_{dc}(T, S)$.

  Similarly, we also have the following three inequalities
  \begin{eqnarray*}
    c_{dc}(G(i, j, k, m), S') \leq  c_{dc}(G(i, j, k, m), S)\\
    c_{dc}(G'(i, j, k, m), S') \leq  c_{dc}(G'(i, j, k, m), S)\\
c_{dc}(G''(m), S') \leq  c_{dc}(G''(m), S)
\end{eqnarray*}
for any $i, j, k, m$.
 Thus, the fact holds. $\Box$
\vspace{1em}

 \noindent  {\bf Fact 2.} In  $S|_{U}$,
   all the leaves $x_i$ must be below one child of
 the root and all the leaves $y_i$ must be below the other child of the root.
 In other words, $S|_{U}=\mbox{LTree}[T_1, T_2]$, where
 $T_1$ is a tree over $x_i$ and some $v_i$s and $T_2$ is a tree over $y_i$s and
some $v_j$s.

\noindent {\bf Proof.}
Assume that  the fact is false. There are $x_i, x_j$ and $y_k$ such that
 $S|_{\{x_i, x_j, y_k\}}= (S|_{U})|_{\{x_i, x_j, y_k\}}= \mbox{LTree}[x_i,
x_j, y_k]$, or there are $y_i, y_j$ and $x_k$ such that
$S|_{\{y_i, y_j, x_k\}}= (S|_{U})|_{\{y_i, y_j, x_k\}}= \mbox{LTree}[y_i,
y_j, x_k]$. If the former is true, then, 
 $$c_{dc}(G_{(i, j, k, m)}, S') \geq 1,\;\; 1\leq m\leq M.$$
This implies that
$$N(N+1)|E|+t\geq  c_{dc}({\cal A}, S') \geq \sum^{M}_{m=1}
c_{dc}(G_{(i, j, k, m)}, S')  = M,$$
  contradicting to the fact that  $M\geq N(N+1)n^2$.
If the latter is true, for any $1\leq m\leq M$,
 $c_{dc}(G'_{(i, j, k, m)}, S') \geq 1$. Again, we have that
$c_{dc}({\cal A}, S')
\geq M$, leading to a contradiction.  $\Box$
\vspace{1em}

  Let $X=\{x_1, x_2, \ldots, x_N\}$ and $Y=\{y_1, y_2, \ldots, y_N\}$.
 Then $S'|_{X}=(S|_{U})|_{X}$ and $S'|_{Y}=(S|_{U})|_{Y}$.
\vspace{1em}

\noindent {\bf Fact 3.} $S'|_{X}=\mbox{LTree}[x_1, x_2, \ldots, x_{N}]$ and
          $S'|_{Y}=\mbox{LTree}[y_1, y_2, \ldots, y_{N}]$.

\noindent {\bf Proof.}  Note that
 $G''_{m}|_{X}=\mbox{LTree}[x_1, x_2, \ldots, x_{N}]$
and  $G''_{m}|_{Y}=\mbox{LTree}[y_1, y_2, \ldots, y_{N}]$ for any
$1\leq m\leq M$.
If the claim is false, then,
 $c_{dc}(G''_{m}, S') \geq 1$ for any $m$ and hence
$$N(N+1)|E|+t\geq  c_{dc}({\cal A}, S') \geq \sum^{M}_{m=1}
c_{dc}(G''_{m}, S')  = M,$$
a contradiction as in the proof of Fact 2. $\Box$
\vspace{1em}

Let  the least common ancestor of $x_i$s and $y_i$s be  $r$ in $S'$.
We have shown that $x_i$s are below one child of $r$, say $r_1$, and
$y_i$'s are below the other child of $r$, say $r_2$. In addition,
$S'|_{X}$ and $S'_{Y}$ are two line trees. \vspace{1em}

\noindent {\bf Fact 4.}  For each edge $e=(v_i, v_j)$ ($i<j$) such that
          $v_i$ and $v_j$ are in the same subtree as $x_i$s or as $y_i$s,
         then
 $$c_{dc}(T_{e1}, S')+ c_{dc}(T_{e2}, S')\geq N(N+1)+1.$$

\noindent  {\bf Proof.} Without loss of generality, we may assume that
         $v_i$ and $v_j$ are  below $r_1$ in the same subtree as $x_i$s.
We consider the
 following cases.

   Case 1. $S|_{X\cup \{v_i, v_j\}}=\mbox{LTree}[x_1, x_2, \cdots, x_k, v_i, x_{k+1},
    \cdots, x_{m}, v_j, v_{m+1}, \cdots, x_{N}]$ for some
 $0\leq k\leq m\leq N$.
   In this case, we have that
     $$ c_{dc}(T_{e1}, S') = \frac{1}{2}N(N-1)+N+1 + \frac{1}{2}(N-k)(N-k-1)$$
and
     $$ c_{dc}(T_{e2}, S') = \frac{1}{2}N(N-1)+k+1 + \frac{1}{2}(N-m)(N-m-1).$$
Hence,
\begin{eqnarray*}
  & & c_{dc}(T_{e1}, S')+ c_{dc}(T_{e2}, S')\\
  & \geq  & N(N-1) + N+1 + \frac{1}{2} [(N-k)(N-k-1)+ 2k+2]\\
  &  \geq &  N(N+1)+1
\end{eqnarray*}
as the minimum value of $(N-k)(N-k-1)+ 2k+2$ is $N$ (reaching at $k=N-2, N-1$).

  Case 2.
    $S|_{X\cup \{v_i, v_j\}}=\mbox{LTree}[x_1, x_2, \cdots, x_k,
 \mbox{LTree}[v_i, v_j],  x_{k+1},
    \cdots, x_{N-1},  x_{N}]$ for some $0\leq k\leq N$.
  We have that
     $$ c_{dc}(T_{e1}, S')=c_{dc}(T_{e2}, S')
  = \frac{1}{2}N(N-1)+k+2 + \frac{1}{2}(N-k)(N-k-1)$$
 and so

\begin{eqnarray*}
 & & c_{dc}(T_{e1}, S')+ c_{dc}(T_{e2}, S')\\
 & \geq & N(N-1) + 2k+4 + (N-k)(N-k-1)\\
 &  \geq &  N(N+1)+2
\end{eqnarray*}
as the minimum value of $2k+(N-k)(N-k-1)$ is  2N-2 (reaching at $k=N-1, N-2$).
The fact is proved. $\Box$
\vspace{1em}

\noindent {\bf Fact 5.}  For each edge $e=(v_i, v_j)$ such that
$v_i$ is  below $r_1$ in the same subtree as $x_i$ and $v_j$ is
below $r_2$  in the subtree as $y_i$s. Then,
 $$c_{dc}(T_{e1}, S')+ c_{dc}(T_{e2}, S')\geq N(N+1).$$

\noindent {\bf Proof.} Let
$$S|_{X\cup \{v_i\}}=\mbox{LTree}[x_1, x_2,\ldots,x_{k}, v_i, x_{k+1},
\ldots, x_{N-1}, x_{N}]$$
and
$$S|_{Y\cup \{v_j\}}=\mbox{LTree}[y_1, y_2,\ldots,y_{m}, v_j, y_{m+1},
\ldots, y_{N-1}, y_{N}].$$

We have that all the internal nodes in $T_{e2}$ are mapped onto the least
common ancestor $r$  of $x_i$s and $y_j$s and thus
  $$ c_{dc}(T_{e2}, S')= N(N+1).$$
Since $c_{dc}(T_{e1}, S')\geq 0$, the fact is proved. $\Box$
\vspace{1em}

 Let $V_1$ denote the subset of leaves $v_i$ below $r_1$ in the same subtree as
$x_i$s and $V_2$ the subset of leaves $v_j$ below $r_2$
 in the same subtree as $y_i$s.
Then $(V_1, V_2)$ is a cut of the graph $\cal G$. Assume there are $p$ cut edges.
Since there are $|E|-p$ non-cut edges,
\begin{eqnarray*}
  & & N(N+1)|E|+t\\
  &= & c_{dc}({\cal A}, S')\\
  & \geq & (|E|-p)N(N+1) + pN(N+1) + (|E|-p)\\
  & = & N(N+1)|E| + |E|-p,
  \end{eqnarray*}
which implies that $p\geq |E|-t$.
This  finishes the proof of Lemma~\ref{lemma2}.
$\Box$

\section{Conclusion}

We conclude this paper by posing two related research problems. In
this paper, we have proved that species tree inference by minimizing
deep coalescences is NP-hard. 
This justifies the effort from different groups in seeking efficient
heuristic methods for the inference problem
\cite{Maddison_SysBiol_06, Tran_PlosCB_09}.
We have also discussed the
relationship of the deep coalescence cost and the gene duplication
cost. Is there any polynomial-time algorithm with constant
approximation ratio for the species tree problem in the deep
coalescence model? Note that  the heuristic method developed by Than
and Nakhleh in \cite{Tran_PlosCB_09} seems to be effective.

In \cite{Fellows_JA_03}, Stege studied the parametric complexity of
the species tree inference by minimizing gene  duplications.
Is is possible to develop efficient algorithm for parametric species tree
inference under the deep coalescence model?

\section*{Acknowledgement}

This work was partially supported by  ARF R146-000-109-112.
A preliminary version of this work was presented in the poster session
of the RECOMB'2000.

\end{document}